\documentclass{osa-article}
\journal{osac}

\usepackage{graphicx}
\usepackage{siunitx}
\usepackage{float}
\usepackage{amsmath}

\begin{document}

% Title
\title{Rapid, accurate, and precise concentration measurements of a methanol-water mixture using Raman spectroscopy}

% Affiliations
\author{Daniel D. Hickstein,\authormark{1*} Russell Goldfarbmuren,\authormark{2}, Jack~Darrah,\authormark{2} Luke Erickson,\authormark{2} and Laura A. Johnson\authormark{3}}

\address{\authormark{1}Kapteyn--Murnane Laboratories Inc., 4775 Walnut St., Suite 102, Boulder, Colorado 80301, U.S.A.\\
\authormark{2}Rebound Technologies, 800 E 64th Ave unit 6, Denver, Colorado, U.S.A\\
\authormark{3}Boulder, Colorado 80305, U.S.A.}

\email{\authormark{*}danhickstein@gmail.com}

% Authors
%\date{\today}

%\setboolean{displaycopyright}{false}

% ABSTRACT
\begin{abstract*}
Here we design, construct, and characterize a compact Raman-spectroscopy-based sensor that measures the concentration of a water--methanol mixture. The sensor measures the concentration with an accuracy of 0.5\% and a precision of 0.2\% with a 1 second measuring time. With longer measurement times, the precision reaches as low as 0.006\%. We characterize the long-term stability of the instrument over an 11-day period of constant measurement, and confirm that systematic drifts are on the level of 0.02\%. We describe methods to improve the sensor performance, providing a path towards accurate, precise, and reliable concentration measurements in harsh environments. This sensor should be adaptable to other water-alcohol mixtures, or other small-molecule liquid mixtures.
\end{abstract*}

%\maketitle

% Section: INTRODUCTION
\section{Introduction}
Raman spectroscopy is a powerful analytical technique that provides detailed information on molecular vibrations. While Raman scattering is typically a weaker effect than directly probing vibrational transitions in the infrared (IR) spectral region ($\sim$3000 to $\sim$20,000 nm), Raman scattering can be accomplished using light in the visible or near-IR region. The ability to use shorter wavelengths can be a significant advantage, because laser sources, spectrographs, and detectors that operate in the visible/NIR region are typically less expensive and more compact than comparable devices in the mid- or long-wave-IR region. Being a convenient and field-deployable technique, Raman spectroscopy has found diverse applications, such as imaging biological samples\cite{camp2014}, detecting trace levels of explosives at a long distance\cite{bremer2013}, and identifying pharmaceutical compounds inside an intact capsule \cite{kim2007}. 

Since a Raman measurement only requires optical access to the sample, and can be completed using relatively inexpensive lasers and spectrometers, it is ideally suited for probing the precise concentrations of mixtures in a laboratory, field, or industrial setting. While Raman spectroscopy has indeed been demonstrated for sensing the concentrations of solutions \cite{rudolph2009, du2015, li2014, kauffmann2012, pandey2017}, liquid mixtures \cite{tiwari2007, holden2003, zehentbauer2012, yu2017}, and gas mixtures \cite{kiefer2008}, the accuracy is typically observed to be on the level of a few percent, the precision is often not quantified, and long-term systematic uncertainties are not investigated. Moreover, demonstrations are typically completed in a laboratory setting and with bulky, expensive equipment. Consequently, it is not clear if a simple, inexpensive Raman spectrometer can be used to measure concentrations with high levels of accuracy, precision, repeatability, reliability, and immunity from environmental perturbations.

% Figure: overview
\begin{figure}
	\begin{center}
	\includegraphics[width=0.9\linewidth]{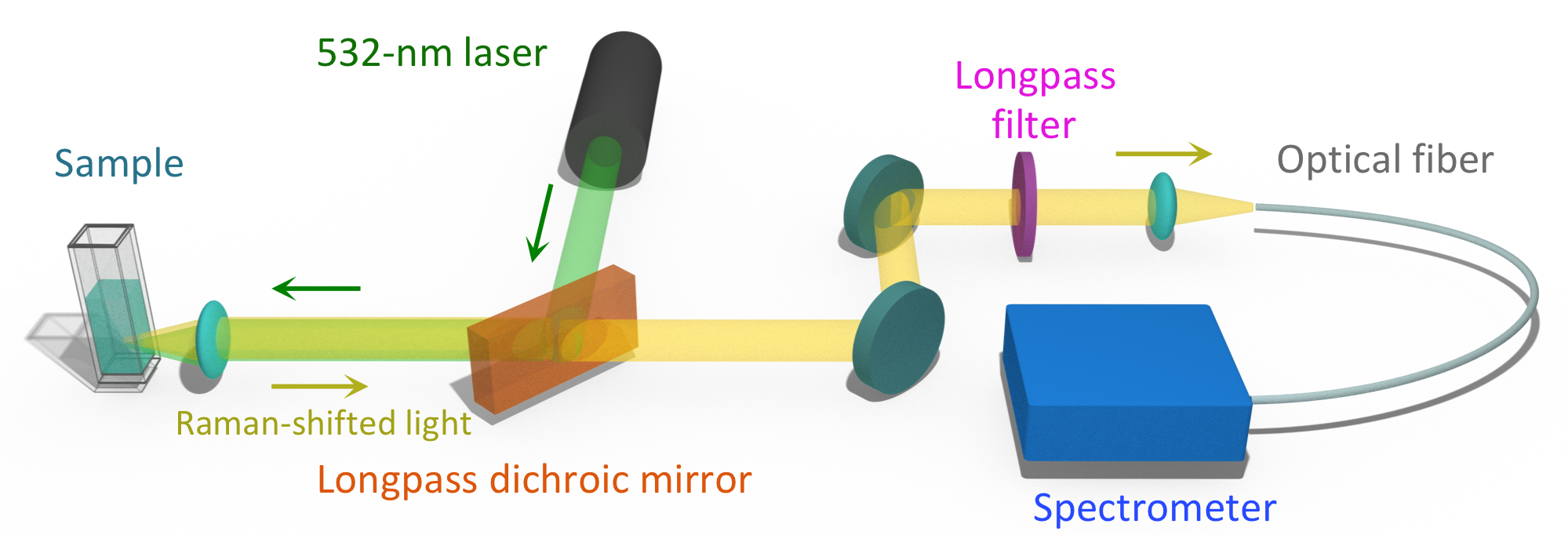}
    \end{center}
	\caption{\label{fig:overview} \textbf{The experimental scheme for the Raman-spectroscopy-based concentration sensor (RCS).} A 532-nm laser is focused into the sample. Some photons experience Stokes Raman scattering, and are emitted in all directions. The back-scattered Raman-shifted photons (yellow) are collected by the same lens, but pass through the longpass dichroic mirror. After passing through a long-pass filter to further reject elastically scattered 532-nm light from the sample, the Raman-shifted light is couped into a fiber which transports it into the grating-based spectrometer.}
\end{figure}

Here we design, construct, and thoroughly characterize a low-cost, low-power-consumption, Raman-spectroscopy-based sensor to precisely measure the concentration of a methanol--water mixture. Our Raman-spectroscopy-based concentration sensor (RCS, Fig.~\ref{fig:overview}) utilizes a excitation wavelength of 532 nm, which allows the use of a low-optical-power (4~mW) laser and an efficient silicon-detector-based spectrometer. By fitting the Raman spectrum of a water--methanol mixture as a linear combination of the spectra of the two constituents, we extract a concentration estimate that is largely immune to variations in laser power, optical alignment, and fluctuations in the absorption of the sample. This simple approach provides an estimate of the methanol concentration to within 3\% accuracy, but with high repeatability. By measuring several mixtures with known concentration, we  prepare an empirical correction curve, which can improve the accuracy of the sensor to better than 0.5\%. We characterize the precision of the sensor on timescales ranging from 0.02 to 600,000 seconds. We find that the sensor provides a precision of 0.2\% with a one-second averaging time, and 0.006\% with a 1000-second averaging time.

% Section: EXPERIMENT
\section{Experiment}
\subsection{Spectrometer design}
The RCS (Fig.~\ref{fig:overview}) is constructed with all commercially available components, with an emphasis on a compact, and durable design. The simple back-reflection geometry\cite{mohr2010} ensures that the strong un-scattered excitation beam is directed away from the detector, relaxing requirements on the long-pass filters. This geometry relies on a dichroic mirror (Thorlabs DMLP567R) to send the laser light to the sample and direct the red-shifted light (generated through spontaneous Stokes Raman scattering) to the spectrometer. The laser light is focused into the sample using a 5-mm-focal length lens, which also collects the scattered light. The sample is contained in a cuvette tube (Thorlabs CV10Q3500) made from 1-mm thickness quartz.

The excitation laser (Thorlabs CPS532) is a compact diode-pumped--solid-state (DPSS) neodymium-doped yttrium aluminium garnet (Nd:YAG), with an internal frequency doubling crystal to provide 4.5~mW (manufacturer's specification) of continuous-wave 532-nm light. We note that this is not a ``single frequency laser'', and changes in the laser frequency and power as a function of time and temperature are expected.

The energy range of Raman shifts that can be measured by our spectrometer includes the important C-H stretch (${\sim} 3000 \,\mathrm{cm^{-1}}$), O-H stretch (${\sim } 3400\,\mathrm{cm^{-1}}$), and C-O stretch (${\sim} 1000\,\mathrm{cm^{-1}}$) regions. The range is limited on the low-energy side by our choice of long-pass filters, and on the high energy side by the range of the spectrometer. A 550-nm long-pass filter (Thorlabs FEL0550) is used to reject elastically (Rayleigh) scattered 532-nm light from reaching the spectrometer. This longpass filter sets a hard cut-off to the observable Raman shift at 615~$\mathrm{cm^{-1}}$. The dichroic mirror has a nominal cut-off wavelength of 567~nm, but, in fact, provides a slow decrease in transmission from 580 to 550~nm, meaning that our Raman spectrometer should exhibit a decreasing response for Raman shifts lower than 1550~$\mathrm{cm^{-1}}$. 

The spectrometer (Ocean Optics USB4000) is a compact grating-based spectrometer with a silicon array-detector, and measured the spectrum at 3647 pixels from 340.2 to 1036.3~nm. This allows us to measure Stokes Raman shifts up to 9150~$\mathrm{cm^{-1}}$ and (if a notch filter were used rather than the long-pass filter) anti-Stokes Raman light up to 10,600~$\mathrm{cm^{-1}}$, a range that is more than sufficient for most applications.

The entire apparatus, including the laser, optics assembly, and spectrometer is housed in a steel box, which reduces sensitivity to ambient light, air currents, vibration, and dust. Beam steering mirrors are held with commercial mirror mounts (Thorlabs) and attached to an optical breadboard (Thorlabs MB612F). 

\begin{figure}
	\begin{center}
	\includegraphics[width=0.8\linewidth]{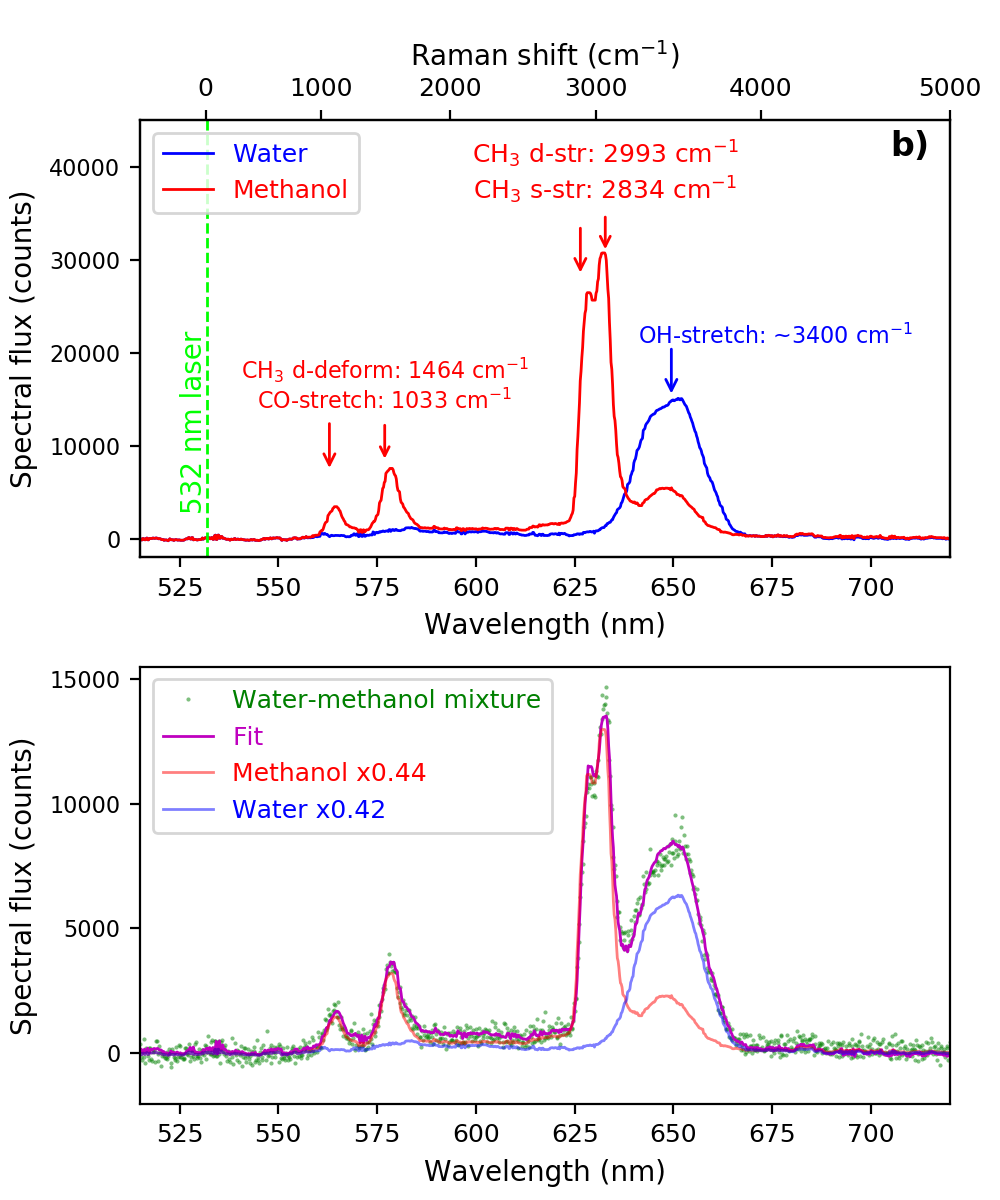}
    \end{center}
	\caption{\label{fig:spectra} a) Raman spectra for methanol pure methanol and pure water samples. These spectra are an average of 100 spectra, each collected for 2 seconds. The various peaks in the methanol spectrum can be assigned to various vibrational transitions listed in the NIST WebBook \cite{jacox2018}. b) The concentration of the mixture of can be estimated by fitting the observed spectrum (green) with a linear combination (magenta) of water (blue) and methanol (red) reference spectra. The coefficients applied to the water and methanol spectra do not sum to one, likely due to fluctuations in the spectral intensity resulting from fluctuations in the laser power. Taking into account the density of methanol, we can extract a weight percent of 44.4\% methanol. Using the empirical correction factor discussed below, we obtain a concentration of 47.3\%, which is the same value as obtained by weighing the water and methanol using a scale.}
\end{figure}

Example spectra for water, methanol, and a mixture of water and methanol are shown in Fig.~\ref{fig:spectra}. For water, a broad peak is seen in the region of 3400~$\mathrm{cm^{-1}}$, which corresponds to the O-H stretching vibration. A very broad peak can also be seen across the 1000 to 2800~$\mathrm{cm^{-1}}$ region, which corresponds to a bending mode. For methanol, narrower, overlapping peaks resulting from C-H stretch are seen near 3000~$\mathrm{cm^{-1}}$, and additional peaks for C-O stretch and a $\mathrm{CH_3}$ bending mode are seen 1400 and 1050~$\mathrm{cm^{-1}}$ respectively. For comparison, the exact frequencies for the Raman transitions listed in the NIST WebBook \cite{jacox2018} are shown in Fig.~\ref{fig:spectra}a.

% Figure: spectra

\subsection{Data processing}
\label{subsec:dataprocessing}
Minimal data processing was required to extract the methanol concentration from the Raman spectrum, and the complete procedure is described here. First, a 9-pixel median filter was applied to remove effects from ``hot pixels'' in the spectrometer CCD and to smooth the acquired spectra. The spectra were clipped to the 600 to 670-nm region, which contains the C-H stretch peak from methanol and the O-H stretch peak from water. We note that we achieved roughly comparable results when using data from the 550 to 670-nm region, but found that expanding the region could increase susceptibility to broadband photoluminescent background that was present in some methanol samples, presumably from impurities.

We fit the spectrum of the methanol-water mixture $f_\mathrm{mix}(\lambda)$ as a linear combination of a pure water reference spectrum $f_\mathrm{w}(\lambda)$ and a pure methanol reference spectrum $f_\mathrm{me}(\lambda)$, according to
\begin{equation}
f_\mathrm{mix}(\lambda) = af_\mathrm{w}(\lambda) + b f_\mathrm{me}(\lambda) + c\lambda + d
\end{equation}
where $a$, $b$, $c$, and $d$ are coefficients determined by the fitting procedure and $\lambda$ is the wavelength. The offset ($d$) and the arbitrary linear term ($c\lambda$) are not strictly necessary, but allow the fitting procedure to be resilient to changes in ambient light levels. The nonlinear fit was completed using the Levenberg--Marquardt algorithm \cite{levenberg1944} as implemented in MINPACK \cite{more1984} and accessed through the \texttt{scipy.optimize.curve\_fit} function in the Scipy Python package\cite{jones2001}. The ``initial guess'' values were $a_0=0.5$, $b=0.5$, $c=0$, and $d=0$. The methanol mass fraction was estimated as 
\begin{equation}
\label{eq:concentration}
w_\mathrm{me} \equiv \frac{m_\mathrm{me}}{m_\mathrm{w} + m_\mathrm{me}} \approx \frac{b\rho_\mathrm{me}}{ a\rho_\mathrm{w} - b\rho_\mathrm{me}},
\end{equation}
where $\rho_\mathrm{w}$ and $\rho_\mathrm{me}$ are the densities of water and methanol. The densities must be included to convert from the Raman measurement, which measures the volume fraction, to mass-fraction, so that we can compare our measurements to the mass-fractions obtained using a scale. In principle, one could simply estimate the methanol mass fraction as $b\rho_\mathrm{me}$, however, by dividing by the total signal in Eq.~\ref{eq:concentration}, we can make the concentration estimate immune to fluctuations in the total spectral intensity, which can be caused by changes in the laser intensity, alignment, or spectrometer response. Fig.~\ref{fig:spectra}b shows an example spectrum of a methanol-water mixture fit as a linear combination of the water and methanol spectra. While we achieve good success with this procedure for our two-component mixtures, we note that more sophisticated algorithms are available to treat multi-component mixtures \cite{gawinkowski2014}.

\subsection{Sample preparation}

Samples of known mass-fraction of methanol and water were prepared using a scale, which had a precision of approximately 1~mg. The total sample volume that was prepared was approximately 1~g. We prepared the samples by adding a volume of methanol, measuring the weight, and then adding a volume of water and measuring the total weight. Consequently, in a situation where we measured 500~mg of methanol and 1000~mg of total weight, the actual mass fraction of methanol might have been $499/1001 \approx 49.85\%$. Thus, we consider the error from the scale to be $\pm0.15\%$.

The samples were weighed into standard 1.5 mL test tubes and shaken vigorously, before being poured into a quartz cuvette tube for analysis. We note that early trials, where the water and methanol were combined directly in the cuvette tube, displayed anomalous results due to insufficient mixing of the water and methanol. For accurate results, thorough mixing is critical.

% Section: RESULTS
\section{Results and Discussion}

% Accuracy
\subsection{Accuracy}

% Figure: accuracy
\begin{figure}[!tb]
	\begin{center}
	\includegraphics[width=0.7\linewidth]{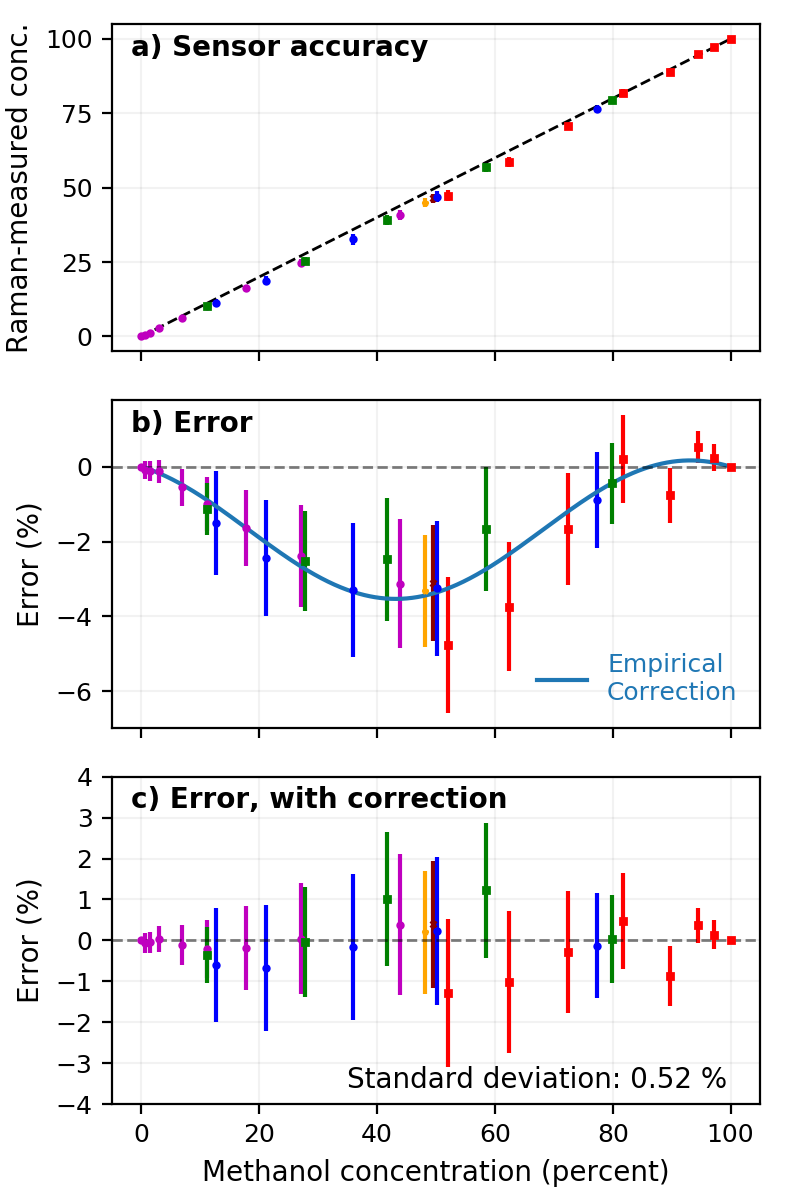}
    \end{center}
	\caption{\label{fig:accuracy} \textbf{The accuracy of the Raman concentration sensor.} a) The extracted concentration of a water-methanol mixture extracted from the Raman spectrum versus the mass fraction measured using a scale. Each point corresponds to a separate measurement, consisting of an average of 10 spectra, each collected using a 1-second exposure time. The various colors indicate independent experiments, conducted on three separate days, demonstrating the repeatability of the measurements. A systematic error underestimate of the methanol concentration is observed, which becomes most pronounced for methanol concentrations near 50\%. b) The difference between the points in panel (a) and the true mass fraction (dashed line in panel (a)) highlight a systematic underestimate of the methanol concentration. The error-bars correspond to two standard errors extracted from the fitting procedure. The line is a simple cosine function that has been fit to the data to model the systematic error. c) Using this empirical correction function, we can correct the observed methanol concentrations and achieve improved agreement between the observed and true methanol concentrations.}
\end{figure}

To assess the accuracy of the RCS, we prepared samples of known weight-percent methanol using a scale, as described above. We then recorded reference spectra of water, methanol, and mixtures of known concentration using a 2-second integration time and averaging 10 spectra. As shown in Fig.~\ref{fig:accuracy}a, the weight-percent methanol recovered from the Raman sensor is true to the real concentration to within a few percent. The agreement is best when nearly pure methanol or nearly pure water is used. But, a systematic error is observed, and when the concentration is close to 50\%  there is a systematic offset of about 3\% (Fig.~\ref{fig:accuracy}b). This accuracy level is similar to the ``detection limit'' for the concentration of a water-toluene mixture reported by Hashimoto et al.\cite{hashimoto2016} using a sophisticated femtosecond laser system.

It is likely that our observed deviation from the true methanol mass fraction is a result of real changes in the mixture. Indeed, it has been shown that water-methanol\cite{holden2003, yu2017, singh2010, nedic2011} and water-ethanol \cite{burikov2010, nedic2011, li2018} mixtures form different water--methanol complexes at different concentrations, and that these complexes display fundamentally different Raman spectra from that of the pure compound. Thus, fundamentally, the spectrum of a water-methanol mixture cannot be perfectly modeled as a linear combination of the spectra of the two components. Thus, any method that makes this assumption, including methods that use the peaks heights, peaks areas, or complete spectral shape (as done here) will experience errors depending on the nature of the complexes that form in the mixture. 

Nevertheless, while it may be difficult to extract an absolute concentration measurement from a Raman spectrum from first-principles, it's possible to empirically correct for the observed deviation, as long as there is a one-to-one relationship between the extracted concentration and the true concentration. In other words, we can prepare mixtures of a known concentration and, as long as the observed concentration monotonically increases with the true concentration, then we can calculate the true concentration from an empirical dataset. For the water-methanol mixtures that we prepare, we observe that our method of fitting the spectra of the mixture as a linear combination does indeed provide a one-to-one mapping to the true methanol concentration. Thus, we can apply an empirical correction based on our recorded data. To obtain the correction function, we fit a cosine function through the data-points in Fig.~\ref{fig:accuracy}b and obtain a satisfactory fit. Using this as an empirical correction, we show that the RCS can provide an estimate of the true methanol concentration with a standard deviation of 0.52\%. 

This behavior was repeatable, and each color in Fig.~\ref{fig:accuracy} represents an independent trial, where new pure-water and pure-methanol reference spectra were collected and fresh mixture samples were prepared. These trials were collected on three separate days, separated by several weeks. Thus, after empirical calibration, the RCS provides repeatable estimates of the methanol concentration with high accuracy. We speculate that the accuracy obtainable with a RCS may be  even better than we report, and we suspect that our estimated accuracy may be limited by our ability to repeatably measure and mix $\sim$1~mL samples. The use of better mixing techniques, a higher precision balance, or larger sample preparation volumes may allow even better accuracy to be realized. As shown in the next sections, the precision and stability of the RCS are sufficient to support measurements better than 0.1\%, and, consequently, the empirically corrected concentration measurements should, in principle, provide the same level of accuracy.

% Precision
\subsection{Precision} \label{subsec:precision}

% Figure: exposure
\begin{figure}
	\begin{center}
	\includegraphics[width=0.8\linewidth]{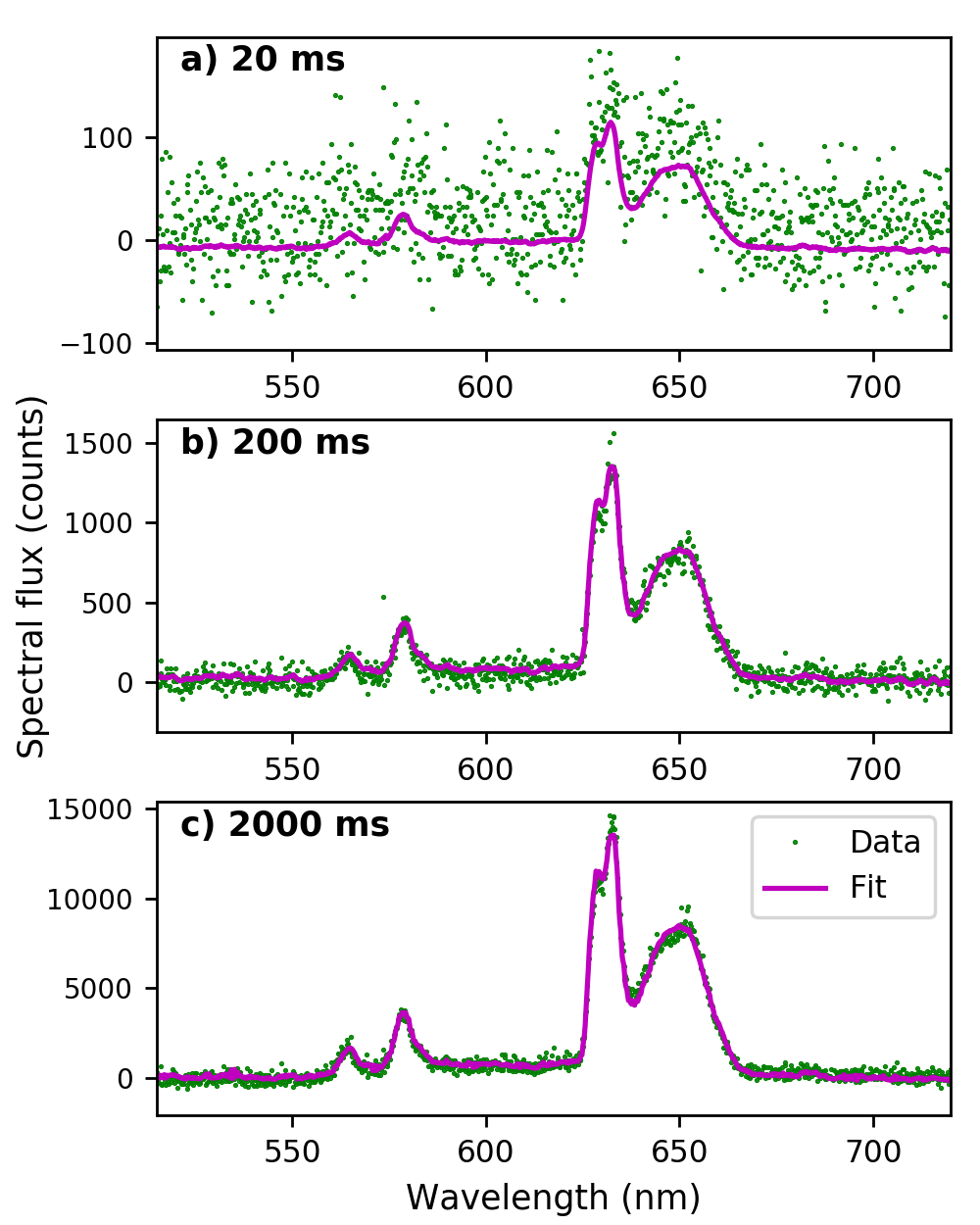}
    \end{center}
	\caption{\label{fig:exposure} The Raman spectra (green dots) show improved signal-to-noise ratio (SNR) as the exposure time is increased. Practically, the exposure time could not be increased indefinitely due to the finite well depth of the spectrometer. Though the 20 ms spectrum exhibits poor SNR, it is still possible to fit the spectrum and obtain an estimate of the methanol concentration.}
\end{figure}

Like many instruments, the precision of the RCS depends on the time available for the measurement. If a longer time is allowable for the measurement, better precision can, in principle, be obtained. The only parameter that can be adjusted in our measurements is the integration time of the spectrometer, which can be equal to or smaller than the total measurement time. For example, if an application requires a concentration measurement every 1 second, then we can envision two ways of performing this measurement. First (Averaging Method~A) we could simply set the exposure time of the spectrometer to 1 second and extract the concentration from the acquired spectrum. Second (Averaging Method~B), we could set the exposure time to 0.1 seconds, collect 10 spectra, extract the concentration from each spectrum, and average the concentrations. We would expect Method~A to provide an advantage in a situation where sensor noise was limiting the precision of the measurement, since it would allow for the highest signal-to-noise ratio for an individual spectrum. Method~B might excel in a situation where other systematic errors dominate.

% Figure: precision
\begin{figure*}
	\begin{center}
	\includegraphics[width=1.1\linewidth]{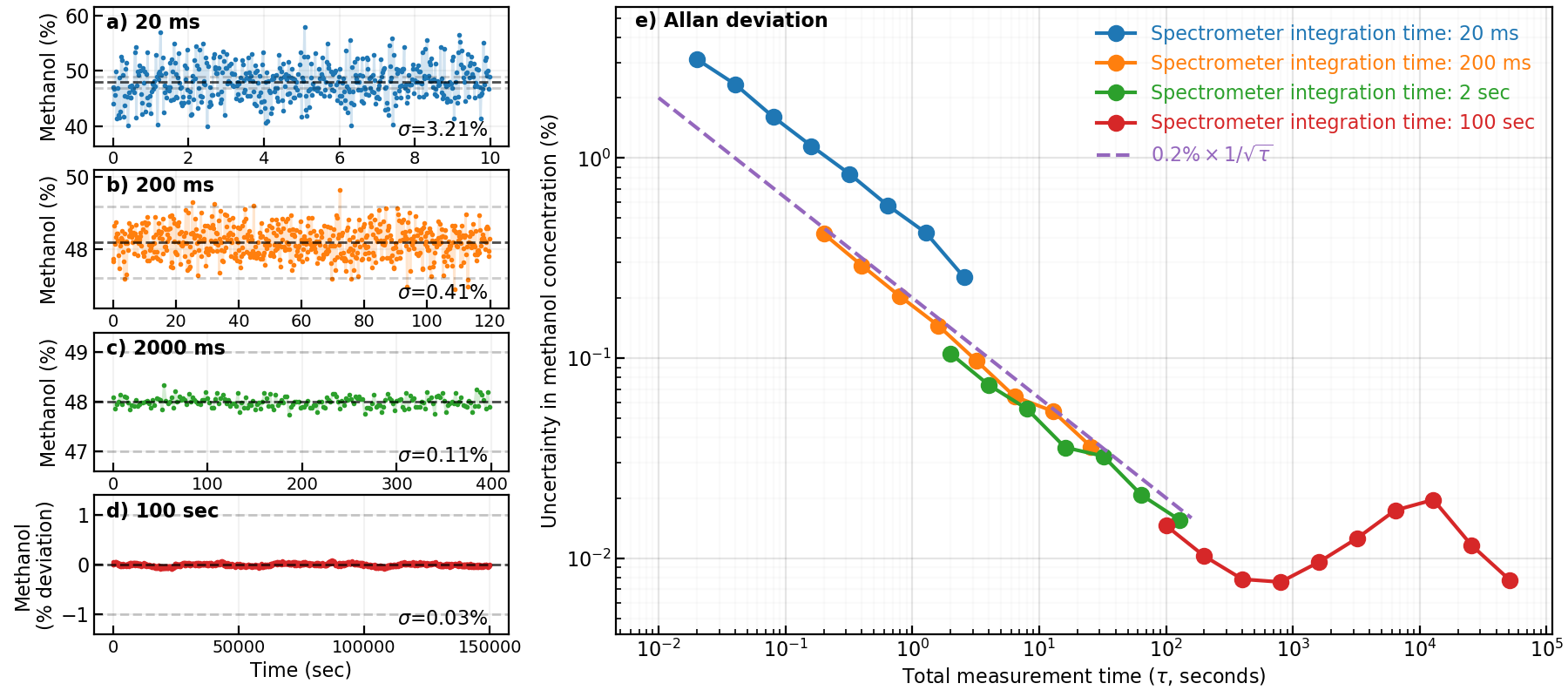}
    \end{center}
	\caption{\label{fig:precision} \textbf{Precision of the Raman-based methanol sensor.}  a--d) The methanol concentration acquired with various measurement times shows improved precision with increasing measurement time. A black dashed line is drawn at the mean concentration, and gray dashed lines are drawn at $\pm$1\% to facilitate comparison. e) The overlapping Allan deviation of each time series (a--d) provides an estimate of the uncertainty in the measurement for a given measurement time. There is a clear trend that the uncertainly decreases with increasing spectrometer integration time, and with the overall measurement time. For the spectrometer integration longer than 200~ms, the precision roughly follows the trend of $0.2\% \times 1/\sqrt{\tau}$. (Note: 100-second measurement time (d) was acquired using an exposure time of 2000~ms and averaging of 50~spectra. Additionally, the data in (d) were corrected for a slow drift, as described in Section~\ref{subsec:longterm}, and (d) presents the deviation from the fit.) }
\end{figure*}

Fig.~\ref{fig:exposure} displays the spectra recorded with various exposure times, and demonstrates that with very low integration times, the Raman features are obscured by detector noise. Consequently, we expect that a longer measurement time will have a double benefit -- not only will more Raman-scattered photons be collected, but a longer integration time can be used, improving the signal-to-noise ratio of each collected spectrum. Indeed, the extracted concentration versus time is quite noisy when the 20~ms integration time is used, and becomes much flatter when a 2000~ms integration time is used (Fig.~\ref{fig:precision}a--d). 

In order to visualize the relationship between measurement time and measurement uncertainty, we employ the overlapping Allan deviation \cite{allan1966, howe1981}, which is the IEEE standard method \cite{ieee2009} for quantifying the instability of frequency references, such as atomic clocks. The Allan deviation is closely related to the standard deviation, but it provides more robust performance in situations that exhibit ``random walk'' noise. Its use here allows for a straightforward analysis of the achievable precision for a given measurement time. Indeed, the overlapping Allan deviation presented in Fig.~\ref{fig:precision}e, enables the comparison of the precision achieved by various exposure times of the spectrometer. For a given exposure time, the uncertainty versus the measurement time ($\tau$) roughly follows the $1/\sqrt{\tau}$ trend would be expected for a signal that randomly varies with white (flat) frequency noise. In addition, the measurement becomes more precise as the spectrometer exposure time is increased. Thus, we conclude that the best precision is offered by setting the spectrometer integration time to be as long as the desired measurement time. In other words, Averaging Method~A appears to be the best option.

The uncertainty of the measurement continues to decrease with increasing measurement time until it reaches a value of 0.008\% at a 400 second measurement time. This behavior suggests that the sensor noise of the spectrometer is the limiting factor for measurement times less than about 400 seconds. At measurement times longer than $\sim$1000 seconds, long-terms drifts begin to exceed the uncertainty of the measurement, and the Allan deviation (Fig.~\ref{fig:precision}e) rises to the 0.02\% level. For times longer than 10,000 seconds, the Allan deviation begins to decrease, indicating the lack of fluctuations on timescales longer than a few hours.

% subsection: longterm
\subsection{Long-term stability} \label{subsec:longterm}
\begin{figure*}
	\begin{center}
	\includegraphics[width=1.1\linewidth]{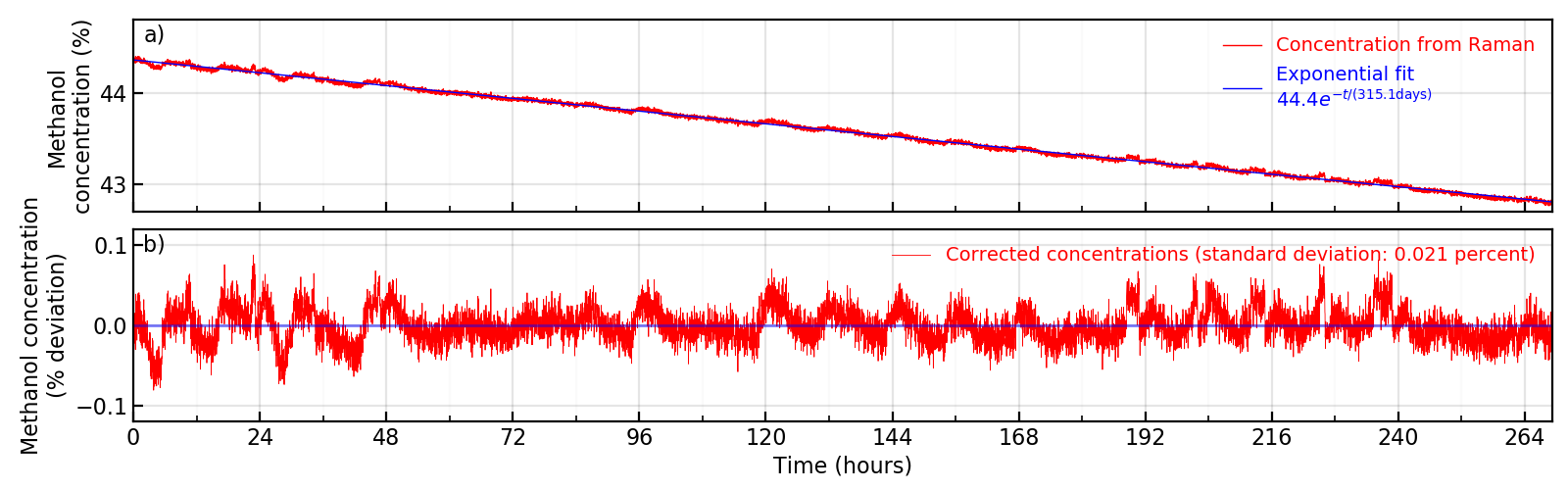}
    \end{center}
	\caption{\label{fig:longterm} \textbf{The concentration of a methanol-water mixture measured constantly over an 11-day period}. a) The concentration shows a slow decrease over the sampling period, which likely results from the evaporation of methanol from the sample. Since the concentration due to evaporation should decay exponentially, an exponential function provides a good fit to the slow trend in the data. b) With the slow drift subtracted, faster drifts, on the several hours timescale can be seen, and correspond to systematic error in the measurement. They are less than 0.1\% for the entire measurement period. We speculate that they are associated with drifts in the laser frequency. Two data-points, corresponding the opening of the box lid to inspect the sample, were omitted. Otherwise, all data-points are plotted, indicating the absence of any spurious readings across the entire measurement period.}
\end{figure*}

In many applications, a sensor would be used to monitor the concentration of a sample for many hours, days, or years. Consequently, a sensor that exhibits minimal drift on long timescales is desirable. To test the stability of our measurement system, we measured the concentration methanol in a water-methanol system over 11.2 days (Fig.~\ref{fig:longterm}). The test was conducted in a non-laboratory environment where twice-daily temperature variations from approximately 12 to 20~C are programmed into the building heating system, with larger local temperature excursions likely. We saved complete spectra continuously over this period, and, to keep the data storage requirements reasonable, we set the spectrometer to average 50 spectra, each with an exposure time of 2~seconds, providing one spectra every 100 seconds. In principle, the concentration could be monitored using a much finer sampling interval with similar results.

% Figure: longterm

Over the 269.4 hour measurement time, the measured concentration of methanol exhibits an overall decrease from 44.3\% to 42.8\% methanol (Fig.~\ref{fig:longterm}a). We attribute this slow decrease to an imperfect seal between the teflon cap and the quartz cuvette, which allowed for some evaporation of the mixture. Since methanol has a higher vapor pressure than water, it evaporates more quickly, leaving the remaining mixture with a lower concentration of methanol. Since the relative evaporation rate of methanol should be proportional to the concentration of methanol, the decay of methanol concentration should be described by $w_\mathrm{me} = w_0 e^{-t /t_\mathrm{evap}}$, where $w_0$ is the initial methanol concentration and $t_\mathrm{evap}$ is the lifetime of the evaporation process. These parameters determined by a similar nonlinear fitting procedure to that described in Section~\ref{subsec:dataprocessing}. By subtracting the exponential function from the data, we can visualize the instrumental fluctuations on the hours-to-days timescale, without the concentration change (Fig.~\ref{fig:longterm}b). The extracted lifetime of the evaporation process $t_\mathrm{evap}$ is 315 days. Consequently, because we subtract an exponential with this time constant, we are not sensitive to systematic drifts on similar timescales. Nevertheless, Fig.~\ref{fig:longterm}b provides insight into systematic errors on the hours and days timescales.

Fig.~\ref{fig:longterm}b shows that the long-term fluctuations of the concentration measurement are below 0.1\% over the entire 11.2 day measurement time, are typically below 0.05\%, and have a standard deviation of 0.021\%. This is below the accuracy of the measurement estimate in Section~\ref{subsec:precision}, and is comparable to the precision achieved with an averaging time of a few-10s of seconds. Consequently, for many applications, the systematic drifts of the measurement will be hidden under the uncertainty of the measurement given by the integration time.

However, even higher precision could be achieved if these systematic fluctuations could be eliminated, so we speculate about the cause of the fluctuations seen in Fig.~\ref{fig:longterm}b. One explanation is that the drifts are real changes in the local methanol concentration probed by the focused laser. During the measurement, we noticed the formation of droplets on the walls of the cuvette (above the level of the liquid). It is likely that the heat from the laser, or thermal gradients within the RCS enclosure, drives an evaporation-condensation cycle within the cuvette, which causes the formation of small droplets on the walls of the cuvette. Since the vapor pressure of methanol is higher than that of water, these droplets would likely have a much higher methanol concentration than the rest of the solution. As the droplets fall back into the solution, the methanol concentration would increase.

However, we find that the fluctuations display a strong Fourier component near $1/12$ hours, suggesting that the fluctuations are related to the twice-daily cycling of the building heating system. There are numerous ways that temperature fluctuations could affect the observed concentration, including changing the alignment of the optical system, affecting the response of the spectrometer, or through a change of the Raman spectrum as a function of temperature. However, we suspect that a change in the wavelength of the 532-nm laser is the most likely cause of the drift. Lasers that have not been specifically stabilized to be ``single frequency'' are well known to exhibit a strong dependence between the wavelength and ambient temperature, and can experience ``mode-hops'' where a small change in temperature results in a sudden change in the laser wavelength \cite{herre1989}. The use of a wavelength-stabilized laser, such as a distributed-feedback (DFB) laser could reduce these drifts and improve the precision of the measurement. A less expensive alternative might be to use a light-emitting diode (LED) as an illumination source \cite{greer2013, schmidt2013}.

% Improvements
\section{Future design improvements}
Above, we have presented a proof-of-principle experiment showing that an compact, accurate, precise, and reliable concentration sensor can be built using relatively inexpensive components. Depending on the application, the sensor described here might provide sufficient performance, or may require further optimization. Here we discuss possible strategies that may be employed for optimizing the size/weight, power consumption, optical access, and robustness to background signals. 

\subsection{Size, weight, and cost}
Currently, the size and weight of the RCS are limited by the use free-space optical components and their associated mounting hardware. The use of fiber-optic components (or miniaturized free-space components) would allow the footprint to be greatly decreased. The cost is currently dominated by the commercial spectrometer. As spectrometer units continue to decrease in price, the use of RCS sensors should become more economical. Alternatively, so-called ``spectral sensors'' provide detection at just a few points across the spectrum, but are very inexpensive, compact, and power efficient. It is possible that recording Raman spectral information at just a few wavelengths is sufficient to estimate the concentration of a simple mixture. 

\subsection{Power consumption}
The specified power consumption of the spectrometer is 250~mA, and the laser is also specified to consume 250~mA. Both require 5~V of DC input. Thus, the nominal power consumption of the instrument is $\sim$2.5 Watts, plus whatever power is required to run the electronics that acquires and processes the data from the spectrometer. In this case, a laptop computer was used to control the spectrometer, greatly increasing the power consumption of the entire system. However, since the processing requirements for recording and processing spectrometer data are quite modest, a micro-controller or a compact single-board computer (for example, a Raspberry Pi, which consumes $\sim$1~Watt) could replace the computer and keep the total power consumption below 4~Watts.

\subsection{Wavelength}
Since the probability for Raman scattering scales as $\lambda^{-4}$, our 532-nm laser provides significantly higher Raman signal per input photon than longer-wavelength lasers. However, the amount of photoluminescent background increases with shorter wavelengths. In our initial tests, we used methanol samples with lower purity, and saw significant photoluminuscent background signals. In many applications, it's likely that the measured mixtures will contain moderate amounts of impurities, which may provide strong photoluminescent background signals. This may complicate the analysis, and prevent the reliable extraction of concentrations. Consequently, for practical applications, the use of a longer wavelength laser may provide the best performance.

% Section: CONCLUSION
\section{Conclusion}
Here we have demonstrated a proof-of-principle device for using Raman spectroscopy to measure the concentration of a mixture of water and methanol. The system consists of all commercial components, including a low-power 532-nm laser and a silicon-detector-based spectrometer. With an empirical correction factor, we can measure the concentration of methanol to better than 0.5$\%$ accuracy and to 0.006\% precision. The instrument can operate over the entire range of methanol concentrations, from 0 to 100\%. We operated the sensor for over 11 days of continuous measurements, and found that the systematic fluctuations remained below 0.1\%. We expect that similar performance should be realized for other water-alcohol mixtures, and likely for any simple liquid mixture where the components have distinct Raman spectral signatures. We have outlined several ways that the design of the concentration sensor may be improved, providing an avenue for the development of field-deployable Raman-based concentration sensors.

\section{Acknowledgments}
We acknowledge helpful discussions with Kyle Schnitzenbaumer, Paul Vallett, David Carlson, Ron Johnson, and Carol Johnson. We thank Kate and Milo Goldfarbmuren for their moral support. 

R.G., J.D., and L.E. are employed by Rebound Technologies, a company that manufactures refrigeration units based on liquid mixtures, including water and methanol. D.D.H works for Kapteyn-Murnane Laboratories, Inc., a company that manufactures laser systems. The authors declare no other competing interests.

\bibliography{raman}

\end{document}